\documentclass[fleqn,twoside]{article}
\usepackage{espcrc2}

\usepackage{graphicx}

\RequirePackage{xspace}
\usepackage{relsize}
\usepackage{color}
\def\babar{\mbox{\slshape B\kern-0.1em{\smaller A}%
           \kern-0.1em B\kern-0.1em{\smaller A\kern-0.2em R}}}

\def\etal    {\textit{et al.}}
\let\semicolon\;
\def\;{\ensuremath{\semicolon}}
\def\CP      {\ensuremath{C\!P}\xspace}
\def\BR      {{\ensuremath{\cal B}}\xspace}
\def\B       {\ensuremath{B}\xspace}
\def\Bbar    {\kern 0.18em\overline{\kern -0.18em B}{}\xspace}
\def\Nbar    {\kern 0.18em\overline{\kern -0.18em N}{}\xspace}

\def\BB      {\ensuremath{B\Bbar}\xspace} 
\def\Bz      {\ensuremath{B^0}\xspace}
\def\Bzb     {\ensuremath{\Bbar^0}\xspace}
\def\BzBzb   {\ensuremath{\Bz {\kern -0.16em \Bzb}}\xspace}
\def\Bu      {\ensuremath{B^+}\xspace}
\def\Bub     {\ensuremath{B^-}\xspace}
\def\Bp      {\ensuremath{\Bu}\xspace}
\def\Bm      {\ensuremath{\Bub}\xspace}
\def\Bpm     {\ensuremath{B^\pm}\xspace}
\def\Bmp     {\ensuremath{B^\mp}\xspace}
\def\BpBm    {\ensuremath{\Bu {\kern -0.16em \Bub}}\xspace}

\def\Y#1S{\ensuremath{\Upsilon{(#1S)}}\xspace}
\def\FourS {\Y4S}
\let\alphatmp\alpha   \def\alpha   {\ensuremath{\alphatmp}\xspace}
\let\betatmp\beta     \def\beta    {\ensuremath{\betatmp}\xspace}
\let\gammatmp\gamma   \def\gamma   {\ensuremath{\gammatmp}\xspace}
\def\ccbar {\ensuremath{c\overline c}\xspace}
\def\qqbar {\ensuremath{q\overline q}\xspace}

\def\stwob{\ensuremath{\sin\! 2 \beta}\xspace}

\def\piz   {\ensuremath{\pi^0}\xspace}
\def\pip   {\ensuremath{\pi^+}\xspace}
\def\pim   {\ensuremath{\pi^-}\xspace}

\def\pipi  {\ensuremath{\pi^+\pi^-}\xspace}
\def\rhoz  {\ensuremath{\rho^0}\xspace}
\def\rhop  {\ensuremath{\rho^+}\xspace}
\def\rhom  {\ensuremath{\rho^-}\xspace}
\def\rhopm {\ensuremath{\rho^\pm}\xspace}
\def\K     {\ensuremath{K}\xspace}
\def\Kp    {\ensuremath{K^+}\xspace}
\def\Km    {\ensuremath{K^-}\xspace}
\def\Kpm   {\ensuremath{K^\pm}\xspace}
\def\Kmp   {\ensuremath{K^\mp}\xspace}

\def\KS    {\ensuremath{K^0_{\scriptscriptstyle S}}\xspace} 
\def\KL    {\ensuremath{K^0_{\scriptscriptstyle L}}\xspace} 
\def\Kz    {\ensuremath{K^0}\xspace}
\def\Kstarz  {\ensuremath{K^{*0}}\xspace}
\def\D       {\ensuremath{D}\xspace}
\def\Dbar    {\kern 0.18em\overline{\kern -0.18em D}{}\xspace}

\def\Dzparst {\ensuremath{D^{(*)0}}\xspace}
\def\Dzbparst{\ensuremath{\Dbar^{(*)0}}\xspace}

\def\DzDzb   {\ensuremath{\Dz {\kern -0.16em \Dzb}}\xspace}

\def\Dz      {\ensuremath{D^0}\xspace}
\def\Dzb     {\ensuremath{\Dbar^0}\xspace}
\def\Dstar   {\ensuremath{D^{*}}\xspace}
\def\Dstarp  {\ensuremath{D^{*+}}\xspace}
\def\Dstarm  {\ensuremath{D^{*-}}\xspace}
\def\jpsi     {\ensuremath{{J\mskip -3mu/\mskip -2mu\psi\mskip 2mu}}\xspace}
\def\psitwos  {\ensuremath{\psi{(2S)}}\xspace}
\def\etac     {\ensuremath{\eta_c}\xspace}
\def\chicone  {\ensuremath{\chi_{c1}}\xspace}
\def\bpsiks   {\ensuremath{\Bz \to \jpsi \KS}\xspace}
\def\bpsikl   {\ensuremath{\Bz \to \jpsi \KL}\xspace}

\def\etaf    {\ensuremath{\eta_f}\xspace}

\def\alphaeff{\ensuremath{\alpha_{eff}}\xspace}

\def\epem    {\ensuremath{e^+e^-}\xspace}

\def\deltat{\ensuremath{{\rm \Delta}t}\xspace}
\def\deltamd{\ensuremath{{\rm \Delta}m_d}\xspace}

\def\bsection#1{\section{\boldmath #1}}
\newcommand{\bsubsection}[1]{\subsection[#1]{\boldmath #1}}

\newcommand{\AmS}{{\protect\the\textfont2
  A\kern-.1667em\lower.5ex\hbox{M}\kern-.125emS}}

\title{CKM-UT Angles: Mixing and \CP violation at the \B Factories}

\author{G. Finocchiaro\address[LNF]{Laboratori Nazionali di Frascati dell'INFN \\
        Via Enrico Fermi 40, \\
        P.O. Box 00044, Frascati, Rome, Italy \\
  },
        on behalf of the \babar\ and Belle Collaborations.}
\begin{document}

\begin{abstract}
We review the experimental status of the angles of the Unitarity Triangle
of the CKM matrix, as measured by the \babar\ and Belle experiments.
\vspace{1pc}
\end{abstract}
\maketitle
\section{Introduction}
The \B Factories have demonstrated since the beginning of this decade that
\CP violation in the \B meson system is consistent with the Standard Model
(SM) description in terms of the complex phase in the three-by-three
Cabibbo-Kobayashi-Maskawa (CKM) matrix\;\cite{ref:CKM}.
With one single phase, the SM predicts clear patterns for quark mixing and
\CP violations, to be satisfied by all processes.
The unitarity relation $V_{ud}V^*_{ub}+V_{cd}V^*_{cb}+V_{td}V^*_{tb}=0$
among the first and third columns of the CKM matrix is represented in
the complex plane by a Unitarity Triangle (UT) with angles
$\alpha=arg[-{V_{td}V^*_{tb}}/{V_{ud}V^*_{ub}}]$,
$\beta=arg[-{V_{cd}V^*_{cb}}/{V_{td}V^*_{tb}}]$,
$\gamma=arg[-{V_{ud}V^*_{ub}}/{V_{cd}V^*_{cb}}]$.
Physics beyond the SM could in general change the
picture; for this reason it is very important
to make many independent measurements to possibly find
inconsistencies of the SM.
\def\tCP{\ensuremath{t_f}\xspace}
\def\ttag{\ensuremath{t_\mathrm{tag}}\xspace}
In the evolution of \BzBzb pairs, we reconstruct the decay of one meson
to final $f$ at proper time \tCP, and identify (tag) its
flavor using information from the other \B meson in the event,
decaying at time \ttag. The time-dependent \CP asymmetry of \Bz(\Bzb) mesons
decaying to final state $f$ can be defined as
$A_{\scriptstyle{CP}}(\deltat)\equiv(N_f-\Nbar_f)/(N_f+\Nbar_f)
=S_f\sin(\deltamd\deltat)-C_f\cos(\deltamd\deltat)$.
Here $\deltat\equiv \tCP-\ttag$, and \deltamd is the mass difference
of the \B meson mass eigenstates.%
\footnote{Some authors, including the Belle
collaboration, use the symbols $\phi_2,\phi_1,\phi_3$ for the angles
$\alpha,\beta,\gamma$, and $A_f=-C_f$ for the parameter describing direct
\CP violation. In the present article we will follow the
$\alpha,\beta,\gamma,C_f$ nomenclature.}. 
The sine term
results from the interference
between direct decay and decay after a $\Bz-\Bzb$ oscillation. A
non-zero cosine term arises from the interference between decay amplitudes
with different weak and strong phases (direct \CP violation) or from \CP
violation in $\Bz-\Bzb$ mixing (the latter is predicted to be small in the
SM and has not been observed to date).

The results discussed in the present paper were obtained by the
\babar\,\cite{ref:babarNIM} and Belle\,\cite{ref:belleNIM} experiments,
respectively located at the PEP-II and KEKB \epem asymmetric-energy \B
factories. Here pairs of \BB mesons are produced almost at rest in the
decay of the \FourS resonance. The separation between their decay vertices
is increased in the laboratory frame due to the boost given by the
asymmetric-energy beams.
The \babar\ experiment has concluded the data taking, collecting a total of
$531\,fb^{-1}$, of which $433\,fb^{-1}$ on the \FourS peak, corresponding
to about $470\times10^6\,$\BB pairs. Belle have logged about $850\,fb^{-1}$
($730\,fb^{-1}$ on the \FourS resonance) as of June 2008. The results
discussed in the present report refer to about $383\times10^6\,$\BB pairs
(\babar) and about $535\times10^6\,$\BB pairs (Belle) unless otherwise
noted.

\begin{figure}[!htb]
  \vspace{9pt}
  \includegraphics*[width=\hsize]{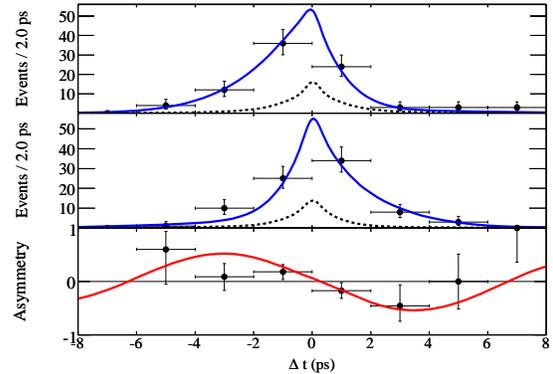}
  \vspace{-20pt}
  \caption{$\Bz\to\jpsi\piz$: \deltat distribution for a sample of
  signal-enriched events tagged as \Bz (top) and \Bzb (middle).
  The bottom plot shows the \deltat asymmetry
  $(N_{\Bz}-N_{\Bzb})/(N_{\Bz}+N_{\Bzb})$. The solid
  curves represent the fitted distributions; the dashed
  line the background contribution.}
  \label{fig:babar_jpsipi0}
\end{figure}

\bsection{Measurements of $\beta$}
\bsubsection{\stwob from $b\to\ccbar s$}
The \B-Factory paradigm of \CP violation measurements is \stwob from
$b\to\ccbar s$ decays. Being dominated by a single decay amplitude, in the
SM with very good accuracy $C_f=0$ and $S_f=-\etaf\stwob$ for these decays,
with \etaf the \CP eigenvalue of the final state.
The latest measurement from \babar~\cite{ref:babar_sin2b_2006},
$\stwob=0.714\pm0.032\pm0.018$, $C=0.049\pm0.022\pm0.017$%
\footnote{Here and in the following, unless otherwise noted, the first
error is statistical and the second one systematic. They are combined if
only one error is given.} includes
modes with $\etaf=-1$ ($\bpsiks$, $\psitwos\KS$, $\etac\KS$,
$\chicone\KS$), with $\etaf=+1$ (\bpsikl), and
$\Bz\to\jpsi\Kstarz(\piz\KS)$, which has no definite \CP parity. 
Belle's latest published measurement\;\cite{ref:belle_sin2b_2006}
concentrates on $\Bz\to\jpsi\Kz$: $\stwob=0.650\pm0.029\pm0.018$,
$C=0.018\pm0.021\pm0.014$. Belle have recently published an updated
measurement in the $\psitwos\KS$ channel\;\cite{ref:belle_psi2S_2008},
based on $657\times10^6\,$\BB pairs. The results,
$\stwob=0.72\pm0.09\pm0.03$ and $C=0.019\pm0.020\pm0.015$, are in good
agreement with the $\Bz\to\jpsi\Kz$ measurement.
\bsubsection{$b\to\ccbar d$ decays}
This class of decays includes both $\Bz\to\jpsi\piz$, whose expected main
contribution is a color-suppressed internal spectator tree diagram, and 
$\Bz\to D^{(*)+}D^{(*)-}$, dominated by a color-allowed tree diagram.
In either case the weak phase of the involved CKM matrix elements
is the same as in $b\to\ccbar s$ decays, and the SM would predict $C=0$
and $S=\stwob$ in the absence of penguin-mediated contributions.
The new \babar\ measurement of the \CP-even $\Bz\to\jpsi\piz$ channel
based on the full dataset of $466\times10^6\,$\BB
pairs\,\cite{ref:babar_jpsipi0_2008} ($S=-1.23\pm0.21\pm0.04$,
$C=-0.20\pm0.19\pm0.03$), constitutes a 
4-sigma evidence for \CP violation in this channel.
The \deltat distribution for \Bz and \Bzb tagged events is shown in
Fig.\,\ref{fig:babar_jpsipi0}. The branching fraction of $\Bz\to\jpsi\piz$
is relevant for constraining possible penguin contributions to the
SU(3)-related channel $\jpsi\Kz$\,\cite{ref:CPS2005}, and is measured to
be $\BR(\Bz\to\jpsi\piz)=(1.69\pm0.14\pm0.07)\times10^{-5}$.
The recent update from Belle\,\cite{ref:belle_jpsipi0_2008} of the
time-dependent measurement of $\Bz\to\jpsi\piz$ ($S=-0.65\pm0.21\pm0.05$,
$C=-0.08\pm0.16\pm0.05$), is quite consistent with \babar's result.

\begin{figure}[!htb]
  \vspace{9pt}
  \centering\includegraphics*[width=0.8\hsize]{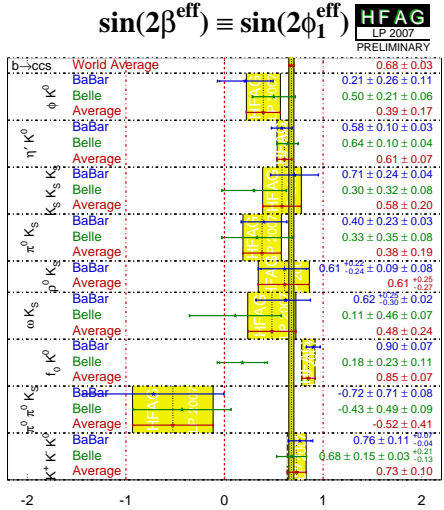}
  \vspace{-20pt}
  \caption{Summary of effective \stwob measurements in
  $b\to s$ decay modes, compared to the world average
  \stwob value in $b\to\ccbar s$.}
  \label{fig:HFAG_sPengS_CP}
\end{figure}

The $\Bz\to\Dstarp\Dstarm$ channel is a Vector-Vector (VV) final state,
which can have $L=0,1,2$ angular momentum and therefore both even and odd
\CP components. It is therefore necessary to measure the \CP-odd fraction
$R_\perp$, and to take into account the dilution due to the admixture.
Belle presented a preliminary update\,\cite{ref:belle_DstDst_2008} of
$R_\perp=0.116\pm0.042\pm0.004$,
and of the \CP-even asymmetry measurement: $S=-0.93\pm0.24\pm0.15$,
$C=-0.16\pm0.13\pm0.02$.
The latest published \babar\
measurement\,\cite{ref:babar_DstDst_2007} found a consistent value of
$R_\perp=0.143\pm0.034\pm0.008$, as well as of the \CP parameters:
$S=-0.66\pm0.19\pm0.04$, $C=-0.02\pm0.11\pm0.02$.
Belle claim\,\cite{ref:belle_DD_2007} 3.2 sigma evidence of direct \CP
violation in $\Bz\to D^+D^-$: $S=-1.13\pm0.37\pm0.09$,
$C=+0.91\pm0.23\pm0.06$. This is unexpected in the SM and not supported by
\babar's measurement\,\cite{ref:babar_DD_2007}, which both in the $D^+D^-$
and in $D^{*\pm}D^\mp$ channels finds \CP asymmetries consistent with the
SM prediction of tree dominance\,\cite{ref:Xing} and therefore with the
result in $b\to\ccbar s$. Since however some new physics models could cause
sizable corrections\,\cite{ref:Grossman}, it is important to keep reducing
experimental uncertainties.
\bsubsection{$b\to\qqbar s$ decays}
The interest of $b\to\qqbar s$ decays has been pointed out for a long
time. The quark transition $b\to s$ is forbidden in the SM at the tree
level, and proceeds dominantly through a penguin diagram
with CKM coefficients proportional to $V_{tb}V^*_{ts}$ and therefore
with the same
weak phase as in \bpsiks decays. Since the tree amplitude is missing, small
effects such as those expected from additional diagrams due to heavy
particles circulating in the loop are in principle more easily
detectable. For this reason these decays are especially sensitive probes of
new physics.
We show in Fig.\,\ref{fig:HFAG_sPengS_CP} a compilation
prepared by the HFAG group\,\cite{ref:HFAG} of available measurement of
$b\to s$ transitions. No recent measurements are available at the time of
the Capri 2008 Workshop, however this is a field of central interest,
where improved experimental precision will hopefully help
clarifying the nature of the small downward shift of $\stwob_{eff}$
observed in most of the $b\to s$ channels respect to the charmonium
reference value.
  \vspace{-20pt}
\bsection{Measurements of $\alpha$}
The angle $\alpha$ is measured with a time-dependent analysis of charmless
decays of neutral \B mesons, $\Bz\to h^+h^-$, with $h=\pi$, $\rho$, $a_1$.
Due to the interplay of tree and penguin diagrams in these channels, the
experiments are actually sensitive to an effective parameter \alphaeff.
As shown in\,\cite{ref:GronauLondon}, one can in principle determine
the shift $\alpha-\alphaeff$ induced by the penguin amplitudes
using the isospin-related decays $\Bz\to h^+ h^-$, $\Bz\to h^0 h^0$ and
$\Bpm\to h^0 h^\pm$. The procedure of measuring the so-called ''isospin
triangles'' requires however rather large datasets, and leaves with up to
eight-fold ambiguities. A relation less stringent, but more accessible with
the current data sample since it does not require to tag the flavor of the
decaying \B also holds\,\cite{ref:GrossmanQuinn}:
$\sin^2(\alpha-\alphaeff)\leq{\BR(\Bz\to h^0 h^0)}/{\BR(\Bpm\to h^0 h^\pm)}$,
which is particularly useful for small values of the numerator.

\begin{figure}[!htb]
  \vspace{9pt}
     \begin{center}
     \includegraphics*[width=0.49\hsize,height=2.7cm]{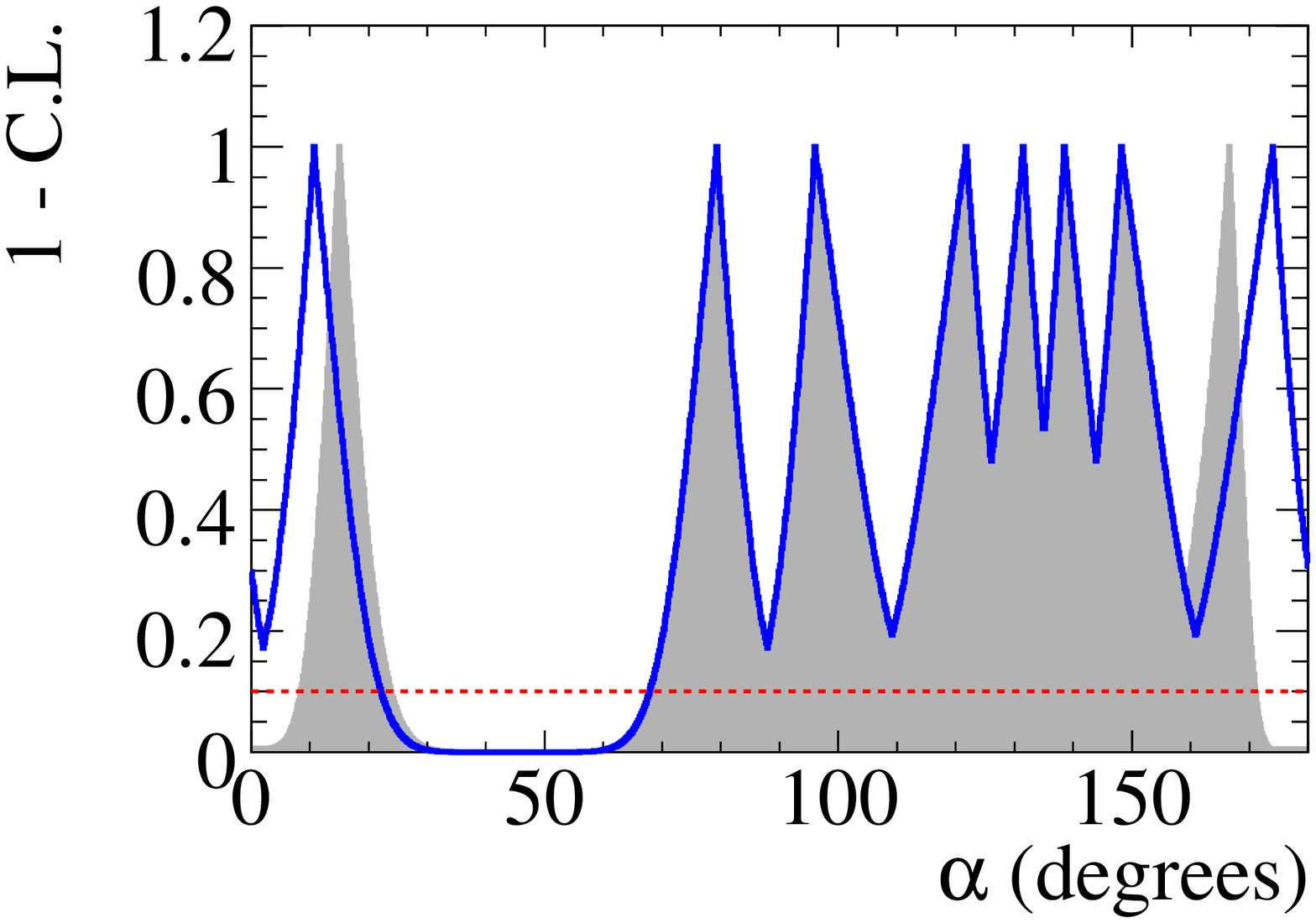}
     \includegraphics*[width=0.49\hsize,height=2.7cm]{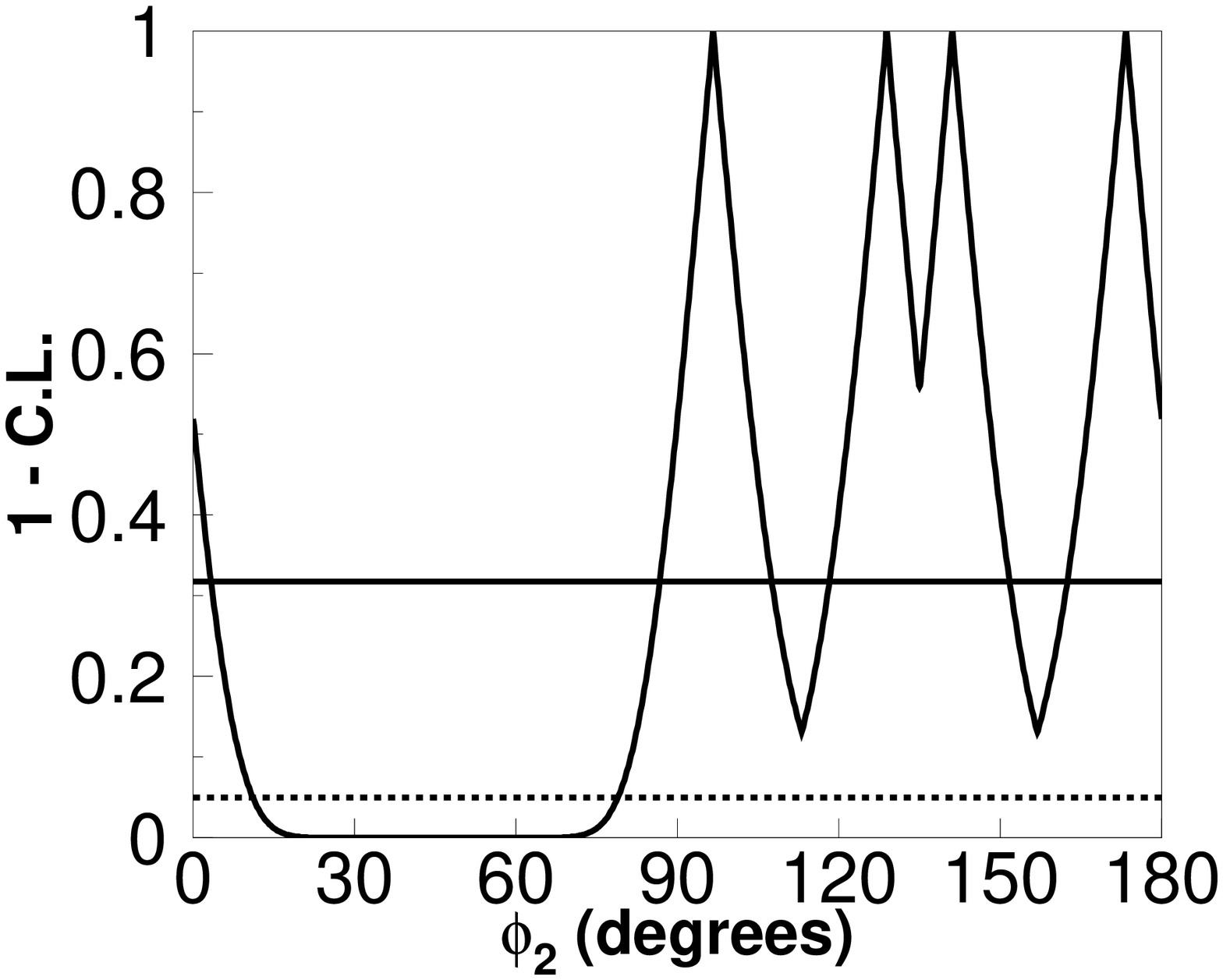}
     \end{center}
  \vspace{-20pt}
    \caption{Constraints on the angle $\alpha$ in $\B\to\pi\pi$ from \babar\
             (left) and Belle (right).}
  \label{fig:babe_alpha_pipi}
\end{figure}

\bsubsection{$\alpha$ from $\B\to\pi\pi$}
This is the ``classic'' channel to measure $\alpha$, with a
well-established evidence for indirect \CP violation by both \B-Factory
experiments, which still show instead a poor (2.1 sigma) agreement on the
$C_{\pi\pi}$ parameter. Both \babar\,\cite{ref:babar_pipi_2007} and
Belle\,\cite{ref:belle_pipi_2007} perform an isospin analysis
to extract $\alpha$, using all the available information ($S_{+-}$,
$C_{+-}$, $C_{00}$, $\BR_{+-}$, $\BR_{+0}$, $\BR_{00}$), shown in
Fig.\,\ref{fig:babe_alpha_pipi}.
One of the allowed solutions ($\alpha=(96^{+11}_{-6})^\circ$ for \babar,
$(\alpha=97\pm11)^\circ$ for Belle) is consistent with the indirect
determination of $\alpha$ in the SM.
\vspace{-20pt}
\bsubsection{$\alpha$ from $\B\to\rho\rho$}
The decay channel $\Bz\to\rhop\rhom$ has the same quark content as \pipi
and can also be used to measure $\alpha$. There are non-trivial
experimental complications due to the presence of two neutral pions in the
final state, with just weak mass constraints from the wide intermediate
resonances. Moreover, $\rhop\rhom$ is a VV state and necessitates in
principle a complete angular analysis to disentangle the
effect of the three possible helicity states. On the other hand, the
branching fraction is about five times larger than $\BR(\Bz\to\pipi)$, and
the state is found to be almost purely longitudinally polarized, so that a
per-event transversity analysis can be avoided and only the longitudinal
\CP parameters need to be determined. There is good agreement between \CP
violation measurement in $\rhop\rhom$ from
\babar\,\cite{ref:babar_rhorho_2007} and
Belle\,\cite{ref:belle_rhorho_2007}. The HFAG average for the longitudinal
components is
$C_{\rhop\rhom}=-0.06\pm0.13$, $S_{\rhop\rhom}=-0.05\pm0.17$.
\babar\,\cite{ref:babar_rho0rho0_2007} presented a preliminary first
time-dependent measurement in the $\Bz\to\rhoz\rhoz$ channel. With
$85\pm28\pm17$ signal events in a sample of $427\times10^6\,\BB$ events,
\babar\ measure $\BR_{\rhoz\rhoz}=(0.84\pm0.29\pm0.17)\times10^{-6}$,
$f_L=0.70\pm0.14\pm0.05$, $S_L=0.5\pm0.9\pm0.2$, $C_L=0.4\pm0.9\pm0.2$.
Consistently, Belle\,\cite{ref:belle_rho0rho0_2008} set the upper limit
$\BR_{\rhoz\rhoz}<1.0\times10^{-6}$ at 90\,\% confidence level (C.L.).

\bsubsection{$\alpha$ from $\Bz\to\rho\pi$}
The third mode used to measure the angle $\alpha$ is
$\Bz\to\pip\pim\piz$. This is not a \CP eigenstate, and four flavor-charge
configurations ($\Bz(\Bzb)\to\rhopm\pi^\mp$) must be considered. The
corresponding isospin analysis is extremely complicated involving
pentagonal relations among the different amplitudes, and cannot be
solved for the 12 unknowns with the present statistics. It was however
pointed out\,\cite{ref:SnyderQuinn} that the variation of the strong phase of
the interfering $\rho$ resonances in the Dalitz plot provides the necessary
degrees of freedom to constrain $\alpha$ with only the irreducible
($\alpha\to\alpha+\pi$) ambiguity. The two \B-Factory experiments have
both performed this analysis. \babar\,\cite{ref:babar_rhopi_2007}
constrain $\alpha=(87^{+45}_{-13})^\circ$;
Belle\,\cite{ref:belle_rhopi_2007} obtain the tighter constraint
$68^\circ<\alpha<95^\circ$ at 68\,\% C.L. for the solution compatible with
the SM.

\bsubsection{$\alpha$ from $\Bz\to a_1\pi$}
The channel $\Bz\to a_1\pi$, which has the same quark content as the
previous modes, has been recently explored in\,\cite{ref:babar_a1pi_2007}.
With a sample of $608\pm52$ signal events, \babar\ adopt a quasi-two-body
approach to obtain a precise measurement of
$\alphaeff=(78.6\pm7.3)^\circ$. Following the proposal
in\,\cite{ref:GronauZupan_2006}, \babar\ are also measuring branching
fractions in the SU(3)-related modes $\B\to a_1K$\,\cite{ref:babar_a1K_2008}
and $\B\to\K_1\pi$ to constrain $|\alpha-\alphaeff|$.

\bsection{Measurements of \gamma}
Several ways have been proposed to measure the angle \gamma at the \B
Factories.
The most effective methods to date exploit direct \CP violation
in $\Bm\to\Dzparst(\Dzbparst)\Km$ decays.
The tree-level decay amplitudes for the $\Bm\to\Dzparst\Km$ and
$\Bm\to\Dzbparst\Km$
transitions differ by a factor $r^{(*)}_B e^{i(\delta^{(*)}_B-\gamma)}$,
where $r^{(*)}_B$ is the magnitude of the $(b\to u/b\to c)$ amplitude
ratio, and $\delta^{(*)}_B$ the strong phase difference.
Estimates of $r^{(*)}_B$ considering CKM and color suppression factors
predict small values, $r^{(*)}_B\simeq0.1\div0.2$.
Different mechanisms have been proposed to obtain interference from
identical \Dz or \Dzb final states. We shall review recent results in the
next subsections.

\begin{figure}[!htb]
  \vspace{9pt}
   \begin{center}
     \includegraphics*[width=0.45\hsize,height=3.5cm]{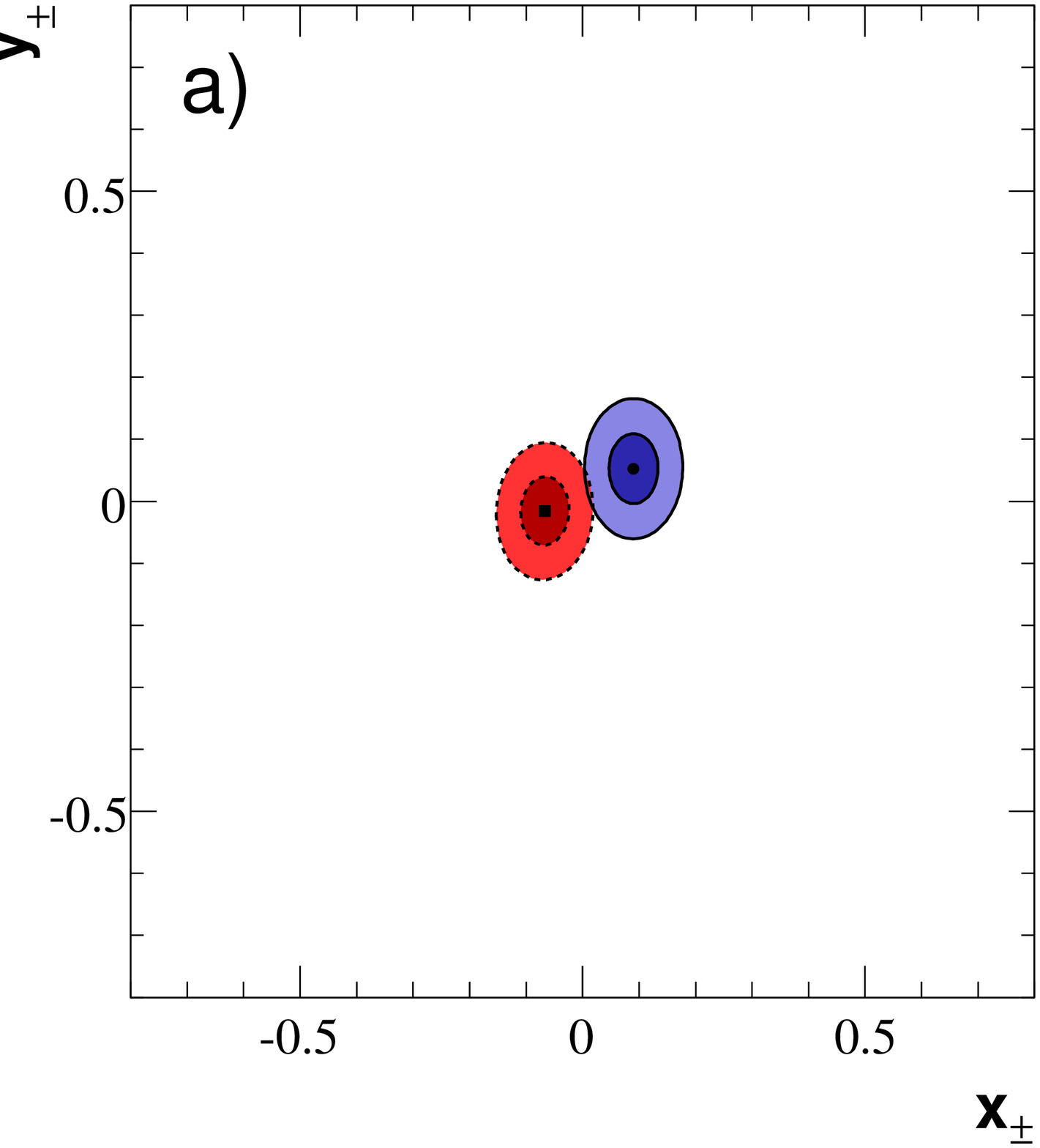}
     \includegraphics*[width=0.45\hsize,height=3.6cm]{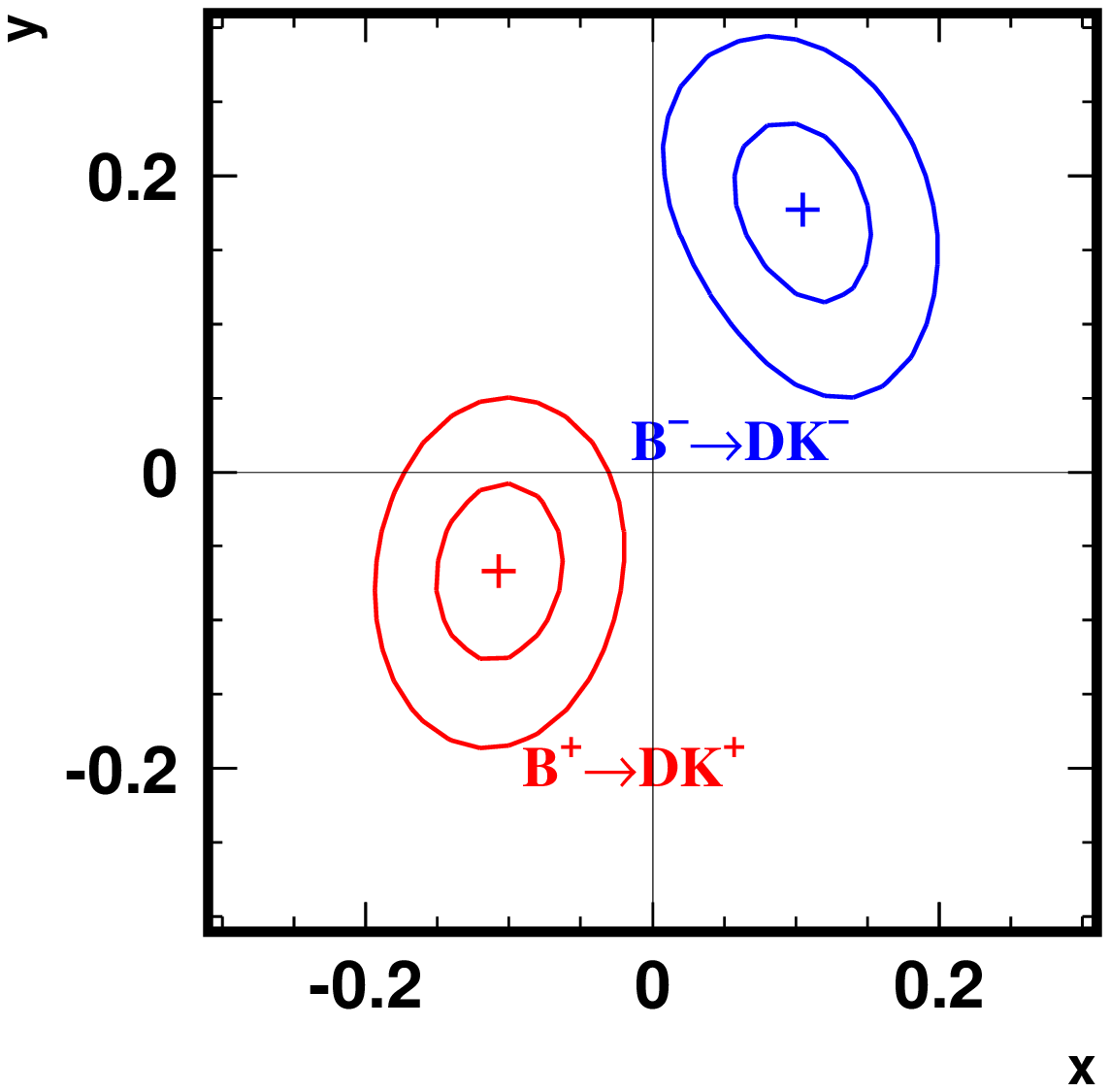}
   \end{center}
\vspace{-25pt}
     \caption{One- and two-sigma 2-dimensional C.L. contours in the
        $(x_\mp,y_\mp)$ plane from \babar\ (left) and Belle (right).}
  \label{fig:babar_gamma_Dalitz_xy}
\end{figure}

\bsubsection{The GLW method}
In the GLW method\,\cite{ref:GLW} neutral \D mesons are reconstructed in
\CP-even ($D_{CP+}$) and \CP-odd ($D_{CP-}$) eigenstates, as well as in
flavor eigenstates (\Dz or \Dzb). The observables
$R_{CP^\pm}\equiv(\BR^-_{CP^\pm}+\BR^+_{CP^\pm})/(\BR^-_{\Dz\Km}+\BR^+_{\Dzb\Kp})/2$
and $A_{CP^\pm}\equiv(\BR^-_{CP^\pm}-\BR^+_{CP^\pm})/(\BR^-_{CP^\pm}+\BR^+_{CP^\pm})$ are measured%
\footnote{We use the compact notation
$\BR^-_{CP^\mp}=\BR(\Bm\to D_{CP\mp}\Km)$,
$\BR^-_{\Dz}=\BR(\Bm\to \Dz\Km)$,
$\BR^-_{\Dzb}=\BR(\Bm\to\Dzb\Km)$, and analogously for the \Bp decays.
}.
These quantities are sensitive to the angle \gamma:
$R_{CP^\pm}=1+r_B^2\pm2r_B\cos\delta_B\cos\gamma$,
$A_{CP^\pm}=\pm2r_B\sin\delta_B\sin\gamma/R_{CP^\pm}$.
\babar\,\cite{ref:babar_GLW_DstK_2008} recently published
updated measurement in $\Bpm\to\Dstar\Kpm$, with $D^{*0}\to\Dz\gamma$,
$\Dz\piz$, and $\Dz(\Dzb)$ reconstructed in \CP-even
($\Kp\Km,\pip\pim$), \CP-odd ($\KS\piz,\KS\omega,\KS\phi$), and
flavor-specific modes ($\Km\pip$), obtaining
$A_{CP^+}=-0.11\pm0.09\pm0.01$,
$A_{CP^-}=+0.06\pm0.10\pm0.02$,
$R_{CP^+}=1.31\pm0.13\pm0.04$,
$R_{CP^-}=1.10\pm0.12\pm0.04$.
The accuracy of these measurements does not allow a determination of
\gamma with the GLW method alone, but contributes improving the overall
precision when combined with the other methods.
\bsubsection{The ADS method}
The idea in the ADS approach\,\cite{ref:ADS} is to select decays with similar
overall amplitudes, in order to maximize the interference and therefore the
sensitivity to \CP asymmetries. This is achieved selecting favored $B\to\D$
decays followed by suppressed \D decays, or viceversa. Analogously to the
GLW case, it is possible to define ratios of branching fractions of
suppressed and favored decays as
$R_{ADS}=\BR_{\B\to\D_{sup}K}/\BR_{\B\to\D_{fav}K}=r^2_D+r^2_B+
2r_Br_D\cos\gamma\cos\delta_B$, and \CP asymmetries as
$A_{ADS}=(\BR_{\Bm\to\D_{sup}\Km}-\BR_{\Bp\to\D_{sup}\Kp})/SUM=
2r_Dr_B\sin\gamma\sin\delta_B/R_{ADS}$.
Belle\,\cite{ref:belle_ADS_2008} recently published an updated
analysis of the suppressed decay chain $\Bm\to D\Km$, $\D\to\Kp\pim$, based
on $657\times10^6\,\BB$ pairs. They do not observe a statistically
significant signal in the suppressed mode, and obtain
$R_{ADS}=(8.0^{+6.3+2.0}_{-5.7-2.8})\times10^{-3}$, 
$A_{ADS}=(-0.13^{+0.97}_{-0.88}\pm0.26)$. These numbers are used to set a
90\,\%\,C.L. upper limit on $r_B<0.19$.
\bsubsection{Dalitz plot method}
Selecting three-body decays of \Dz and \Dzb such as $\KS h^+h^-$
($h=\pi,K$), the Dalitz plot distribution depends on the interference of
Cabibbo allowed, doubly-Cabibbo suppressed and \CP eigenstate decay
amplitudes. Neglecting mixing and \CP violation in the \DzDzb meson system,
the amplitude for $\Bmp\to\D[\KS h^+h^-]\Kpm$ can be written as
${\cal A}^{(*)}_\mp(m^2_-,m^2_+) \propto {\cal A}_{D\mp} +
\lambda r^{(*)}_B e^{i(\delta^{(*)}_B\mp\gamma)}{\cal A}_{D\pm}$,
where $m^2_-$ and $m^2_+$ are the squared invariant masses of the $\KS h^-$
and $\KS h^+$ combinations respectively,
$\lambda=-1$ for $\Dz{}^*\to\gamma\Dz$ and $=1$ otherwise,
and ${\cal A}_{D+}$ (${\cal A}_{D-}$) are the amplitudes of the
$\Dz(\Dzb)\to\KS h^+h^-$ decay, described with a detailed model
involving several intermediate resonances and
extracted from large control samples of
flavor-tagged $\Dstarp\to\Dz\pip$ decays produced in \ccbar events.
The 'cartesian' variables
$x^{(*)}_\mp=r^{(*)}_B\cos(\delta^{(*)}_B\mp\gamma)$,
$y^{(*)}_\mp=r^{(*)}_B\sin(\delta^{(*)}_B\mp\gamma)$ are used by the
experiments to avoid the bias due to $r^{(*)}_B$ being positive definite.
In their preliminary work\,\cite{ref:belle_gamma_Dalitz_2008} based on 
$657\times10^6\,\BB$ pairs, the Belle collaboration reconstruct the decays
$\Bmp\to\D^{(*)0}\Kmp$, with $\D^{*0}\to\Dz\piz$ and $\Dz\to\KS\pip\pim$,
producing new and more precise results for the
$(x,y)^{(*)}_\mp$ parameters.
With a statistical procedure they find 
$r_B=0.16\pm0.04$, $r^*_B=0.21\pm0.08$ and $\gamma=(76^{+12}_{-13})^\circ$.
\babar\ is also publishing an updated
result\,\cite{ref:babar_gamma_Dalitz_2008},
based on $383\times10^6\,\BB$ pairs. In addition to the modes used
by Belle, \babar\ also reconstruct the decays $\D^{*0}\to\Dz\gamma$,
$\Bmp\to\Dz\K^{*\mp}[\KS\pi^\mp]$, and $\Dz\to\KS\Kp\Km$.
As an illustration, results for $(x,y)_\mp$ in the $\Bmp\to\Dz\Kmp$ mode from
\babar\ and Belle are shown in Fig.\,\ref{fig:babar_gamma_Dalitz_xy}.
\vspace{-25pt}
\vspace{-20pt}
Thanks to the larger number of reconstructed channels and to a better
analysis efficiency, \babar\ determines
the $(x,y)^{(*)}_\mp$ parameters with the same accuracy as Belle despite
the smaller data sample, finding 3 sigma evidence of \CP violation.
However, the \babar\ data favor smaller $r_B$ values ($r_B=0.086\pm0.035$,
$r^*_B=0.135\pm0.051$, $\kappa r_s=0.163^{+0.088}_{-0.105}$)\footnote{The
amplitude ratio in $\Bmp\to\Dz\K^{*\mp}$ events is described
by $\kappa r_s$, with $\kappa$ taking into account non-resonant $\KS\pi^\mp$
contributions .},  and thus a larger error for \gamma
($\gamma=(76^{+23}_{-24})^\circ$).

\section{Summary and outlook}
The \B Factories have established \CP violation in several
\B decays, and measured \stwob in charmonium decays with precision better than
4\%. All \beta measurements in many different channels are consistent. Some
channels, such as the penguin-dominated $b\to s$ modes are particularly
promising because they are especially sensitive to heavy virtual states.

The angle \alpha is being studied in charmless $\Bz\to\pi\pi$, $\rho\rho$
and $\rho\pi$ transitions. The first measurements of the $\Bz\to\rhoz\rhoz$
decay confirm the indication that the effect of penguin amplitudes is
relatively small in $\rho\rho$ decays, which in fact yield the most
stringent constraints on \alpha. New channels such as $\Bz\to a_1\pi$ and
SU(3)-related decays are being studied, and will hopefully contribute to
improve the determination of \alpha, which will eventually be limited by
penguin pollution.

A precise measurement of the angle \gamma, simply unthinkable at the
beginning of the \B-Factory era, is now a reality thanks to the large
accumulated statistics and the number of \B decays sensitive to this
angle. Several new measurements of $\Bmp\to\Dz\Kmp$ 
transitions have appeared recently, and strong evidence for direct \CP
violation in these decays is building up. The Dalitz method in particular
provides the most stringent constraints to date.

All measurements of CKM angles are at present statistically limited, and
will therefore become more precise in the near future, when the \babar\
collaboration analyze their full dataset, and Belle continue to accumulate
new data.


\begin{thebibliography}{99}
\bibitem{ref:CKM}
 N. Cabibbo, Phys. Rev. Lett. \textbf{10}, 531 (1963);\\
 M. Kobayashi and T. Maskawa, Prog. Th. Phys. \textbf{49}, 652 (1973).
\bibitem{ref:babarNIM}
  B. Aubert \etal, the \babar\ Collaboration,\\
  Nucl. Instr. and Methods {\bf A479}, 1 (2002).
\bibitem{ref:belleNIM}
  A. Abashian \etal, the Belle Collaboration,\\
  Nucl. Instr. and Methods {\bf A479}, 117 (2002).
\bibitem{ref:babar_sin2b_2006}
  B. Aubert \etal, the \babar\ Collaboration,\\
  Phys. Rev. Lett. {\bf 99}, 171803 (2007).
\bibitem{ref:belle_sin2b_2006}
  K.-F. Chen \etal, the Belle Collaboration,\\
  Phys. Rev. Lett. {\bf 98}, 031802 (2007).
\bibitem{ref:belle_psi2S_2008}
  K.-F. Chen \etal, the Belle Collaboration,\\
  Phys. Rev. {\bf D77}, 091103 (2008).
\bibitem{ref:babar_jpsipi0_2008}
  B. Aubert \etal, the \babar\ Collaboration,\\
  Phys. Rev. Lett. {\bf 101}, 021801 (2008).
\bibitem{ref:CPS2005}
  M. Ciuchini, M. Pierini and L. Silvestrini,\\
  Phys. Rev. Lett. {\bf 95}, 221804 (2005).
\bibitem{ref:belle_jpsipi0_2008}
  S. E. Lee \etal, the Belle Collaboration,\\
  Phys. Rev. {\bf D77}, 0711011 (2008).
\bibitem{ref:belle_DstDst_2008}
  The Belle Collaboration,\\
  Moriond 2008 preliminary, unpublished.
\bibitem{ref:babar_DstDst_2007}
  B. Aubert \etal, the \babar\ Collaboration,\\
  Phys. Rev. {\bf D76}, 111102 (2007).
\bibitem{ref:belle_DD_2007}
  S. Fratina \etal, the Belle Collaboration,\\
  Phys. Rev. Lett. {\bf 98}, 221802 (2007).
\bibitem{ref:babar_DD_2007}
  B. Aubert \etal, the \babar\ Collaboration,\\
  Phys. Rev. Lett. {\bf 98}, 071801 (2007).
\bibitem{ref:Xing}
  X. Y. Pham and Z.-Z. Xing,
  Phys. Lett. {\bf B 458}, 375 (1999); 
  Z.-Z. Xing, Phys. Rev. {\bf D61}, 014010 (2000).
\bibitem{ref:Grossman}
  Y. Grossman and M. P. Worah,\\
  Phys. Lett. {\bf B395}, 241 (1997).
\bibitem{ref:HFAG}
  The Heavy Flavor Averaging Group, \\
  \texttt{http:$\!$/$\!$/www.slac.stanford.edu/xorg/hfag}
\bibitem{ref:GronauLondon}
  M. Gronau and D. London,\\
  Phys. Rev. Lett. {\bf 65}, 3381 (1990).
\bibitem{ref:GrossmanQuinn}
  Y. Grossman \etal,\\
  Phys. Rev. {\bf D58}, 017504 (1998).
\bibitem{ref:babar_pipi_2007}
  B. Aubert \etal, the \babar\ Collaboration,\\
  Phys. Rev. {\bf D76}, 091102 (2007).
\bibitem{ref:belle_pipi_2007}
  H. Ishino \etal, the Belle Collaboration,\\
  Phys. Rev. Lett. {\bf 98}, 211801 (2007).
\bibitem{ref:babar_rhorho_2007}
  B. Aubert \etal, the \babar\ Collaboration,\\
  Phys. Rev. {\bf D76}, 052007 (2007).
\bibitem{ref:belle_rhorho_2007}
  K. Abe \etal, the Belle Collaboration,\\
  Phys. Rev. {\bf D76}, 011104 (2007).
\bibitem{ref:babar_rho0rho0_2007}
  B. Aubert \etal, the \babar\ Collaboration,\\
  arXiv:0708.1630 [hep-ex].
\bibitem{ref:belle_rho0rho0_2008}
  The Belle Collaboration,\\
  La Thuile 2008 preliminary, unpublished.
\bibitem{ref:SnyderQuinn}
  H.R. Quinn and A.E.Snyder,\\
  Phys. Rev. {\bf D48}, 2139 (1993).
\bibitem{ref:babar_rhopi_2007}
  B. Aubert \etal, the \babar\ Collaboration,\\
  Phys. Rev. {\bf D76}, 012004 (2007).
\bibitem{ref:belle_rhopi_2007}
  A. Kusaka \etal, the Belle Collaboration,\\
  Phys. Rev. Lett. {\bf 98}, 221602 (2007).
\bibitem{ref:babar_a1pi_2007}
  B. Aubert \etal, the \babar\ Collaboration,\\
  Phys. Rev. Lett. {\bf 98}, 181803 (2007).
\bibitem{ref:GronauZupan_2006}
  M. Gronau and J. Zupan,\\
  Phys. Rev. {\bf D73}, 057502 (2006).
\bibitem{ref:babar_a1K_2008}
  B. Aubert \etal, the \babar\ Collaboration,\\
  Phys. Rev. Lett. {\bf 100}, 051803 (2008).
\bibitem{ref:GLW}
  M. Gronau and D. London, Phys. Lett. {\bf B253}, 483 (1991);
  M. Gronau and D. Wyler, Phys. Lett. {\bf B265}, 172 (1991).
\bibitem{ref:babar_GLW_DstK_2008}
  B. Aubert \etal, the \babar\ Collaboration,\\
  arXiv:0807.2408v1 [hep-ex].
\bibitem{ref:ADS}
  D. Atwood, I. Dunietz, and A. Soni,
  Phys. Rev. Lett. {\bf 78}, 3257 (1997);
  Phys. Rev. {\bf D 63}, 036005 (2001).
\bibitem{ref:belle_ADS_2008}
  Y. Horii \etal, the Belle Collaboration,\\
  arXiv:0804.2063v1 [hep-ex].
\bibitem{ref:belle_gamma_Dalitz_2008}
  K. Abe \etal, the Belle Collaboration,\\
  arXiv:0803.3375 [hep-ex].
\bibitem{ref:babar_gamma_Dalitz_2008}
  B. Aubert \etal, the \babar\ Collaboration,\\
  Phys. Rev. {\bf D 78}, 034023 (2008).
\end{thebibliography}
\end{document}